\documentclass[useAMS,usenatbib]{mn2e}

\usepackage{defs}

\usepackage{graphicx}

\def\lsim{\hbox{ \rlap{\raise 0.425ex\hbox{$<$}}\lower 0.65ex\hbox{$\sim$} }}
\def\gsim{\hbox{ \rlap{\raise 0.425ex\hbox{$>$}}\lower 0.65ex\hbox{$\sim$} }}

\def\opeqn{\begin{equation}}
\def\cleqn{\end{equation}}
\def\opeqnno{\begin{equation*}}
\def\cleqnno{\end{equation*}}

\def\dfrac{\displaystyle\frac}
\def\dint{\displaystyle\int}

\pdfoutput = 1

\title[Puzzles in Time Delay and Fermat Principle in Gravitational Lensing]
{Puzzles in Time Delay and Fermat Principle in Gravitational Lensing}
\author[Puzzles in Time Delay and Fermat Principle in Gravitational Lensing]{Sun Hong Rhie\thanks{E-mail: srhie@nd.edu} \\
Physics Department, University of Notre Dame, Notre Dame, 46556, USA.}

\begin{document}

\date{SUBMITTED $14^{\rm th}$ March 2011}

\pagerange{\pageref{firstpage}--\pageref{lastpage}} \pubyear{2011}

\maketitle

\label{firstpage}

\begin{abstract}
The current standard time delay formula (CSTD) in gravitational lensing and its claimed relation to the lens equation through Fermat's principle (least time principle) have been puzzling to the author for some time. We find that the  so-called geometric path difference term of the CSTD is an error, and it causes a double counting of the correct time delay. We examined the deflection angle and the time delay of a photon trajectory in the Schwarzschild metric that allows exact perturbative calculations in the gravitational parameter $GM$ in two coordinate systems -- the standard Schwarzschild coordinate system and the isotropic Schwarzschild coordinate system.  We identify a coordinate dependent term in the time delay which becomes irrelevant for the arrival time difference of two images. It deems necessary to sort out unambiguously what is what we measure. We calculate the second order corrections for the deflection angle and time delay. The CSTD does generate correct lens equations including multiple scattering lens equations under the variations and may be best understood as a generating function. It is presently unclear what the significance is.  We call to reanalyze the existing strong lensing data with time delays.

\end{abstract}



\section{Introduction}

\citet{shapiro} championed the time delay of the
radar echo off of the inner planets as a fourth test of Einstein general relativity.
The so-called classic three tests were suggested by Einstein and they are Mercury precession, 
astrometric shifts of the background stars
due to the gravitational lensing by the Sun, and gravitational redshifts \citep{redshift59}. 
In the same year, \citet{refsdal-delay} suggested to measure the time delay difference
(or arrival time difference) between two images to use in conjunction with other measurables
(position and flux of the images) of supernovae
to determine the Hubble constant and mass of the lens galaxies. 
The time delay measurement is a central issue in strong gravitational lensing where multiple
images are detected, especially because the time data can help break the degeneracy of
the lens fitting models. There have been monitoring campaigns of multiple image quasars
in radio and optical \citep{kochanek08, coles08}.  
With large high cadence survey telescopes in plan
(LSST, WFIRST, EUCLID) time delay measurements will become a routine business, and
it is expected to see many multiply imaged supernovae \citep{kirkby}. 

The time delay function (2d scalar function) is also known to generate the lens equation 
(2d vector equation) through the stationarity
or extremum condition (commonly referred to as the Fermat's principle) for single scattering
lenses \citep{schneider85, pipe} and multiple scattering lenses withal \citep{BN86}. The single
scattering lens equations are well known from direct derivations of the photon path in the 
Schwarzschild metric and its generalizations in the effectively Newtonian gravitational field. 
The multiple scattering lens equation can also be derived directly by looking at the photon 
path in 3-space \citep{RCB} and it confirms the validity of the variational method.

The current standard time delay function is made of the
geometric path difference term and gravitational potential term. 
The so-called geometric path difference term of the CSTD is essentially a quadratic function
of the position difference between the source and image, and we have been
puzzled by it for some time  even though
it is widely used unsuspected and being
incorporated into pipeline codes for systematic studies of gravitational lensing
(\citet{kochanek08}, \cite{coles08}, \citet{pipe}, and references therein). 
We examine the CSTD
for a Schwarzschild black hole lens which has the virtue of allowing exact calculations
of the time delays and find the CSTD wrong. It effectively doubles the correct time delay. 
We trace the origin of the current standard time delay formula form to \citet{CK75} 
(CK75 hereafter)
and find that the so-called geometric path difference term suffers from mistakes. 
CK75 uses the so-called isotropic metric (in the linear approximation). Thus
we discuss the deflection angle and time delay of a photon path in the Schwarzschild metric
in two coordinate systems --  the standard Schwarzschild coordinates and the isotropic
Schwarzschild coordinates, where the latter can be obtained from the former (or vice versa)
by changing the radial coordinates. In the linear order in $r_s \equiv 2GM$ where $M$ is the
mass of the Schwarzschild black hole, the radial coordinates only 
differ by a constant: $r = u + r_s/2$ where $r$ and $u$ are the standard and isotropic
radial coordinates respectively. We will see that the 
two metric forms (both asymptotically flat)
result in the same time delay difference for two images. 
At the same time, it leaves a question as to what is the time the observer measures.
We examine the deflection angle and find that the difference between the coordinate
systems is effectively of the second order. In other words, the deflection angles 
measured in the two coordinate systems are the same while the time delays are 
different. 
 We calculate the second order corrections to the deflection angle and time delay. 
 We conclude in section \ref{sec:conclusion}.

\section{The Current Standard Time Delay Formula}
\label{sec:current}

The current standard time delay formula used in the lensing community 
can be found, for example,  in eq.(A1) of \citet{pipe} which is 
reproduced here for convenience.
\begin{eqnarray}
 {ct(\theta_I; \theta_s) } =  \cr
 (1+z_L)\left( \dfrac{D_L^2}{D}\dfrac{(\theta_I - \theta_s)^2}{2} 
- \dfrac{4G}{c^2}\dint \Sigma(\theta^\prime) \ln|\theta_I - \theta^\prime| d^2\theta^\prime \right).
\label{eqTD}
\end{eqnarray}
The first term is referred to as the time delay due to the geometric path length difference
between the deflected and undeflected paths, and the second term is referred to as the
time delay due to the gravitational potential acting on the photon along the path.

For a Schwarzschild black hole of mass $M$,
eq.(\ref{eqTD}) becomes
\opeqn
ct(\theta_I; \theta_s)  
= (1+z_L)\left( \dfrac{D_L^2}{D}\dfrac{(\theta_I - \theta_s)^2}{2} 
- \dfrac{4GM}{c^2}  \ln|\theta_I - \theta_L|  \right)
\label{eqTDsingle}
\cleqn
where $D_L$ is the distance (from the observer) to the lens and $D$ is the reduced distance.
$\theta$ is the angular position variable, and the subscripts $s, L, I$ denote source, lens, and image
respectively. $z_L$ is the redshift of the lens.
($c$ is the speed of light, which will be set to be 1 hereafter: $c=1$.)
We may say that the time function is made of a ``quadratic term" and a logarithmic term.
The stationary condition
 with respect to the variation of the image position, $dt/d\theta=0$,  generates the lens equation.
 Since $\theta$ is a two-dimensional angular variable in general, we take it as a vector,
 and the resulting equation is the well-known single lens equation.
 \opeqn
 0=  (\theta_I - \theta_s)
- \dfrac{\theta_E^2 \, (\theta_I - \theta_L)}{ |\theta_I - \theta_L|^2} ; \quad \theta_E^2 \equiv \dfrac{4GMD}{D_L^2}
\label{eqSingle}
 \cleqn
 where $\theta_E$ is the angular Einstein radius.
This stationarity condition  is commonly 
referred to as Fermat's principle: an image forms such that the time is an extremum
(even though Fermat's principle for Snell's law invokes the notion of the {\it least time}.)
The single point lens equation (\ref{eqSingle}) has two images whose positions are collinear as is well known.
\opeqn
 \theta_{I \pm}= \dfrac{\theta_s}{2} \left(1 \pm \left(1 + \dfrac{4\theta_E^2}{\theta_s^2} \right)^{1/2} \right)
\cleqn
 $\theta_L$ has been set to zero by translating the coordinate system.
The arrival time difference ($t(\theta_-) - t(\theta_+)$: $\theta_-$ is the position of the dimmer image
and arrives later) due to the quadratic term and the logarithmic term are 
\begin{eqnarray}
 \Delta t_{12} (quad) = (1+z_L) 2GM s(s^2+4)^{1/2}; \quad s^2 \equiv \dfrac{\theta_s^2}{\theta_E^2}  \\
 \Delta t_{12} (log) = (1+z_L) 8GM \ln \left(\dfrac{s}{2} +(\dfrac{s^2}{4}+1)^{1/2} \right).
 \label{eqDeltaT}
\end{eqnarray}
For $s\lsim 1$, the both become $\Delta t_{images} \approx  (1+z_L) 4GM s$. 
The total time delay between two images would be $\approx  (1+z_L) 8GM s$.

\subsection{Refsdal's Time Delay}

\citet{refsdal-delay} shows that the arrival time difference of the two images of a spherical
galaxy lens is $\approx 8GM s$ in eq.(6-refsdal). The time delay formula was derived in 
  eq.(30-refsdal) of  \citet{refsdal}, and we may write it  as follows.
\opeqn
 \Delta t_{12} = \dint_0^s \dfrac{D_s}{D_2} (\theta_1 - \theta_2) ds
  = \dfrac{D_s}{D_2}\dfrac{R_E^2}{D_1} \dint_0^s (s^2+4)^{1/2} ds
  \label{eqRefsdalDef}
\cleqn
where $D_s = D_1 + D_s$ and $s$ is the dimensionless source position variable.
Refsdal didn't present the exact integral, but it is easy to do.
\opeqn
\Delta t_{12} = \dfrac{R_E^2}{D} \left(\dfrac{1}{2} (b_+^2 - b_+^{-2}) + \ln b_+ \right); \ \ 
b_+ \equiv \dfrac{s}{2} +(\dfrac{s^2}{4}+1)^{1/2}.
\cleqn
$b_+$ is nothing but the position of the positive image (in units of the Einstein ring radius).
The negative image position $b_- = s-b_+$ and $b_- = b_+^{-1}$. 
Thus,
\opeqn
\Delta t_{12} = \dfrac{R_E^2}{D} \left(\dfrac{1}{2} ((b_- -s)^2 - (b_+ -s)^2) 
          - \ln \left| \dfrac{b_-}{b_+} \right| \right).
\label{eqRefsdalTD}
\cleqn
It is a surprise; it is exactly the CSTD at $z_L=0$!!
Unfortunately, we have no clue how to understand the definition of the time delay in 
eq.(\ref{eqRefsdalDef}): how integrating the angular separation of the two images
as the source moves from $s=0$ (where there is no arrival time difference between
the two images) to $s=s$ is supposed to correspond to the time delay. 
Refsdal doesn't explain it. It must have been self-evident to him. 
In short, we don't know how he got it wrong.

\begin{figure*}
\includegraphics[trim=2mm 12mm 2mm 18mm, clip, width=13.5cm]{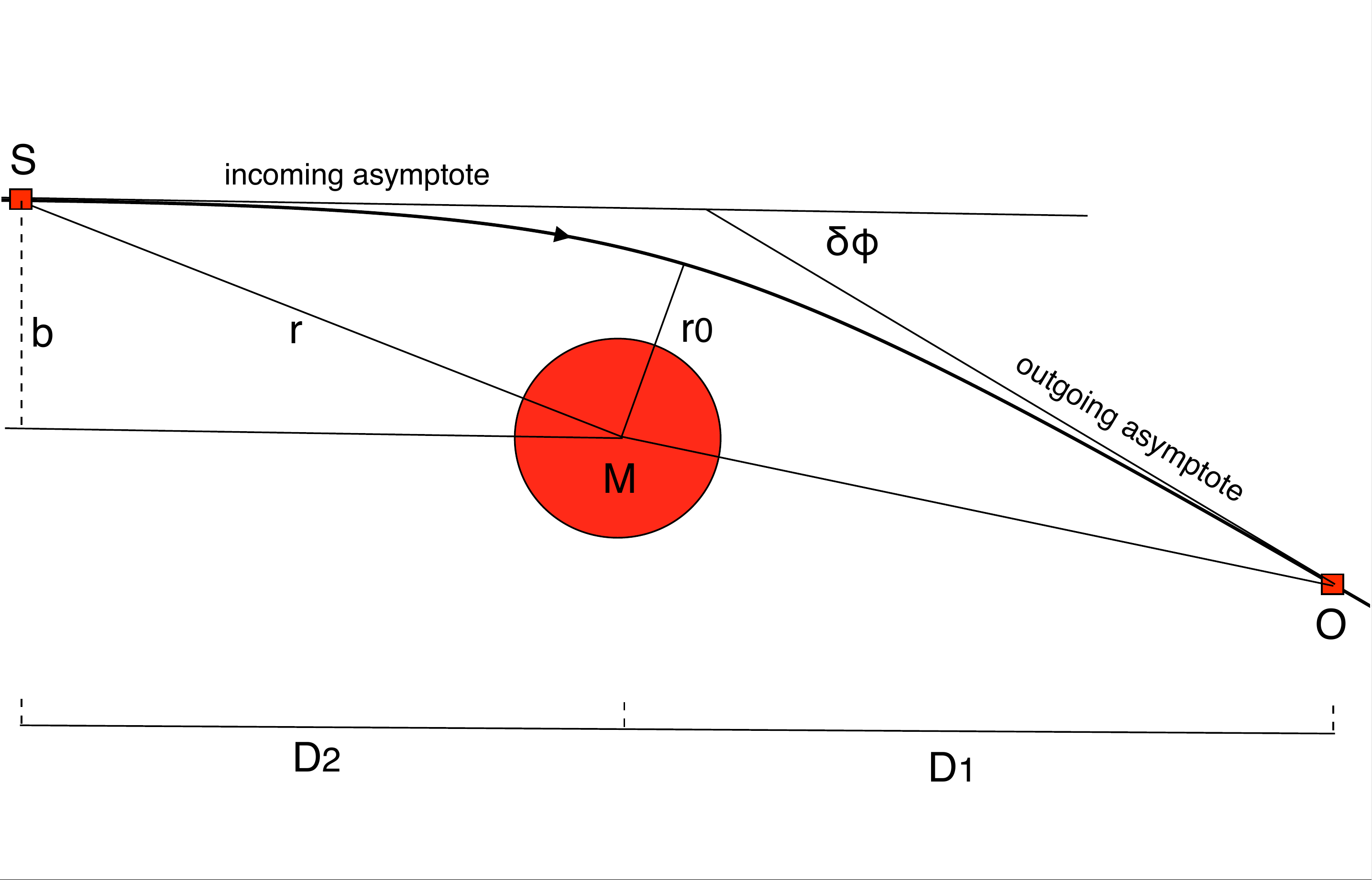}
\caption{Depiction of a photon path (arrowed thick curve), from a source $S$ to an observer $O$,
passing by a spherically symmetric mass $M$ 
in Schwarzschild coordinates. $r_0$ is the distance of the closest approach, $\delta\phi$ is the 
scattering angle, and $b$ is the impact distance. In the small scattering approximation relevant
in most of the astronomical practices, $\delta\phi << 1$, $b\approx r_0$ and $\sqrt{r^2-r_0^2}$
is the path length of the photon from $S$ to the center of the mass $M$ in the absence of the gravitational 
interaction.} 
\label{fig:path}
\end{figure*}

\subsection{Simply Looking at Fig.\ref{fig:path}}  
\label{sec:simply}

 Figure \ref{fig:path} shows a deflected photon path in $(r, \phi)$ scattering plane.
If we look at fig.\ref{fig:path}, the path difference between the actual photon path 
and the would-be straight line is $\Delta\approx r_0 \delta\phi  = 2 r_s$. It is a constant
irrelevantly of the path. 

 We can try to do better. Let's assign viewing angles
by the observer of the image, source, and lens ($M$) as $\alpha$, $\beta$, and 
$\gamma$ respectively, then we may estimate it as $\Delta \approx r_1(\alpha-\gamma)\delta\phi$.
Or, we may estimate it a bit more carefully by looking at the outgoing and incoming paths
separately:  $\Delta_1 \approx r_1 (\alpha - \gamma)^2$ and
 $\Delta_2 \approx r_2 (r_1/r_2)^2 (\alpha - \gamma)^2$. Then
\opeqn
 \Delta = \Delta_1 + \Delta_2 \approx \dfrac{D_1^2}{D} (\alpha - \gamma)^2.
 \label{eqDelta}
\cleqn
The path length difference is larger for the brighter image. In other words, the arrival time
difference of the two images will be opposite to what is implied in the ``quadratic term."
The consequence is that the arrival time difference between the two images will vanish
because the magnitudes of the ``quadratic term" and logarithmic term are practically
equal as we have seen in the previous section.
The lens equation (the relation between the source and image positions) can be found easily.
\opeqn
\alpha -\beta = \dfrac{D}{D_1^2}\delta\phi = \dfrac{R_E^2}{D_1^3 (\alpha-\gamma)}.
\cleqn 
Now we can express the geometric path difference $\Delta$ in terms of $(\alpha-\beta)$.
\opeqn
 \Delta \approx \dfrac{2 r_s R_E^2}{D_1^2 (\alpha-\beta)^2}.
 \label{eqDeltaSource}
\cleqn
It may be best called an ``inverse quadratic term" than ``quadratic term."

\subsection{Gott and Gunn,  Cooke and Kantowski, and Schneider's Fermat Principle}

\citet{GG74} wrote down without explanation the quadratic part of Refsdal's time delay 
in eq.(\ref{eqRefsdalTD}) as the correct time delay between two images. It is hard
to know the underlying reasoning.

\citet{CK75} (CK75 hereafter) clearly lays out a reasoning for their calculation
and hence is of interest here. CK75 interpreted the ``quadratic term" of Gott and Gunn 
as due to the geometric path difference
and purported to ``complete" the time delay formula leading to the CSTD for an 
arbitrary mass distribution where the weak gravity approximation (linear in $GM$) holds. 
 CK75 doesn't mention that the CSTD was obtained by Refsdal for a spherically symmetric
 mass even though they cite Refsdal's 1964a and 1964b. It is possible that CK75 didn't 
 know it, and they don't mention what might be the underlying reason of Refsdal's
 definition of the time delay perhaps because of it. CK75 does offer their own reasoning
 for the ``quadratic term" as has been stated already and lays out a rule for the linear 
 calculation. The ``quadratic term" they obtain is of the second order in $GM$, hence it
 is a violation of their first order approximation assumption. 
 Their calculation of the ``quadratic term" amounts to a confusion where $\gamma$
 in eq.(\ref{eqDelta}) is replaced by $\beta$, which led to the ``quadratic form" 
 instead of the ``inverse quadratic form." The difference of factor 2 comes from 
 the slight difference in the definition of the path difference, (See Fig.2 of CK75.),
 which doesn't concern us especially because the ``quadratic term" will be shown
 to be irrelevant to the time delay.
  
 \citet{schneider85} ``improved" CK75's CSTD and introduced a variational principle
 to derive the lens equation from the CSTD time delay. Schneider called it Fermat's
 Principle. Once the CSTD is found invalid as a measure of the flight time of a photon, 
 the notion of Fermat's principle may be best abandoned.

\section{Time Delay due to a Spherically Symmetric Mass}
 
 \subsection{The Standard Schwarzschild Metric}§

 The deflection angle of a photon trajectory 
 due to a spherically symmetric mass $M$ and its flight duration can be calculated exactly by solving
 the equation of motion in the following standard Schwarzschild metric. 
\opeqn
ds^2 = -B(r)dt^2 + A(r) dr^2 + r^2 d\Omega^2  ; \ \   A(r)^{-1} = B(r) = 1-\dfrac{2GM}{r}.
\label{eqSchwarzschild}
\cleqn
The equation of motion can be derived from the variational principle,
$ 0 = \delta \int ds $.
\opeqn
  0 = \dfrac{d^2x^\mu}{dp^2}  + \Gamma^\mu_{\alpha\beta} 
   \dfrac{dx^\alpha}{dp}\dfrac{dx^\beta}{dp}
\cleqn
where $p$ is the line parameter and  $ \Gamma^\mu_{\alpha\beta}$ is the Christoffel symbol 
that is a function of the linear derivatives of the metric components.
\opeqn
  \Gamma^\mu_{\alpha\beta} = \dfrac {1}{2}g^{\mu\lambda}(\partial_\mu g_{\nu\lambda}
   \partial_\nu g_{\mu\lambda} -  \partial_\lambda g_{\mu\nu} )
\cleqn
where $g^{\mu\lambda}$ is the inverse metric components: $g^{\mu\lambda}g_{\lambda\alpha} = \delta^\mu_\alpha$.
Because of the spherical symmetry, the photon path lies on a plane, say $\theta=\pi/2$, 
and the equations involve three variables: $t$, $r$, and $\phi$. 
\opeqn
 \dfrac{dt}{dp} = \dfrac{1}{B} ; \ \ \
 \dfrac{d\phi}{dp} = \dfrac{J}{r^2}; \ \ \
 A \left(\dfrac{dr}{dp}\right)^2 = \dfrac{1}{B} - \dfrac{J^2}{r^2}.
\cleqn
where $J$ is the angular momentum. From the equations, one can get
 $d\phi = g(r) dr$ or $dt = f(r) dr$. 
The scattering (or deflection) angle is obtained by integrating $d\phi = g(r) dr$.
\opeqn
\int d\phi = \left( \dint_{r_2}^{r_0} +\dint_{r_0}^{r_1} \right ) 
           \dfrac{\sqrt{A}/r \ \ dr }{ \sqrt{ \dfrac{B_0 r^2}{B r_0^2}-1 }  }
\label{eqAngleIntegral}
\cleqn 
 The integration $\int d\phi = \int g(r) dr$ doesn't relent to a nice formula unlike in 
Newtonian but can be done easily 
 by expanding in $GM$ (weak gravity approximation), 
which is suitable for astrophysical purposes.
The scattering angle comes out to be $\delta\phi = 4GM/r_0$ as is well known, and the 
weak gravity approximation is also commonly called the small angle approximation. 
$r_0$ is the distance of the closest approach of the photon path to the mass $M$ 
as is shown in figure \ref{fig:path}. 
 In order to obtain the 
the Schwarzschild coordinate time duration it takes for the photon to travel from the source
to the observer, one can integrate $dt = f(r) dr$ in the same small angle approximation.
\opeqn
\int dt = \left( \dint_{r_2}^{r_0} +\dint_{r_0}^{r_1} \right ) 
           \dfrac{\sqrt{A/B} \ \ dr }{ \sqrt{ 1-\dfrac{B r_0^2}{B_0 r^2} }  }
\label{eqTimeIntegral}
\cleqn
where $B_0 = B(r_0)$.
The duration either from source
 at $r = r_2$ to $r_0$ or from $r_0$ to the observer at $r = r_1$ is given as follows.
 (See, for exmaple,  eq.(8.7.4) of \citet{weinberg}; the speed of light $c=1$.)
\begin{eqnarray}
t(r, r_0)  = 
 \sqrt{r^2-r_0^2}  
+ 2 GM \ln \left(\dfrac{r+\sqrt{r^2-r_0^2}}{r_0}\right) \cr
          +  GM\left(\dfrac{r-r_0}{r+r_0}\right)^{1/2} + {\cal O}({r_s^2}/{r_0^2}).
\label{eqHalfTime}
\end{eqnarray}
where $r_s = 2GM$ is the Schwarzschild radius. The Shapiro time delay formula of the
radar echo eq.(1-shapiro) of \citet{shapiro} can be obtained from eq.(\ref{eqHalfTime}).

Note that the first term does not depend on the mass $M$ and should be the time the photon
takes to travel from the source to the center of the mass $M$ when the gravitation is
turned off. It can be directly calculated from eq.(\ref{eqTimeIntegral}) by setting $A=B=1$
and should also be intuitively evident from figure \ref{fig:path}.
The proper time duration measured by the observer is obtained by multiplying the 
time dilation factor $B(r)^{1/2}$ in eq.(\ref{eqSchwarzschild}) to the Schwarzschild coordinate time 
(\eg, eq.(6.3.46) of \citet{wald}). 
Thus, the total proper time delay due to the general relativity in the linear order in $r_s/r_0$ is  
\begin{eqnarray}
\Delta \tau/r_s = \ln \left(\dfrac{r_2+\sqrt{r_2^2-r_0^2}} {r_0}\right) 
   + \ln \left(\dfrac{r_1+\sqrt{r_1^2-r_0^2}} {r_0}\right) \cr
          + \frac{1}{2} \left( \sqrt{\dfrac{r_2-r_0}{r_2+r_0} } 
          + \sqrt{ \dfrac{r_1-r_0}{r_1+r_0} } \right)
          -  \dfrac{\sqrt{r_2^2-r_0^2}+\sqrt{r_1^2-r_0^2} }{2 r_1} 
\label{eqProperTime}
\end{eqnarray}
where $r_2$ and $r_1$ are the radial positions of the source and observer respectively. 
Now define the distances from the observer to the lens and the source along the horizontal line
$D_1$ and $D_2$ as shown in figure \ref{fig:path}. Using
\opeqn
  r_j^2 \approx D_j^2 + r_0^2  \quad {\rm where} \ \  D_j >> r_0,
\cleqn
we get
 \opeqn
 \Delta \tau/r_s = \ln \dfrac{4D_1D_2}{r_0^2} + \dfrac{D_1-D_2}{2D_1} - \dfrac{r_0}{2D} + {\cal O}(r_0^2/D_j^2)
 \label{eqTimeDelay}
 \cleqn
where $D$ is the reduced distance. ($D_1$ is used in place of $D_L$ 
in section \ref{sec:current} for handy manipulation of the index.) 
If we ignore the third term for now because it is usually very small, the proper time delay between two images, 
1 and 2, depends 
only on the first term.
\opeqn
\Delta t_{12} \equiv \Delta t_1 - \Delta t_2 =  - 2 r_s  \ln (r_{01} /  r_{02}).
\label{eqTimeDelayImage}
\cleqn
This is exactly the same as the arrival time difference due to the logarithmic term of the time
function in eq.(\ref{eqTDsingle}) with $z_L=0$. (Here the expansion 
of the universe is ignored.) The true time delay formula doesn't have
a ``quadratic term."  If we recall that $\Delta t(quad) \approx \Delta t(log)$
when the source is within the Einstein radius from the lens (where the dim image is not too dim), 
the current standard time delay is about twice the true time delay. It is not a small difference
and it makes it urgent to reanalyze the time delay data. Considering that the time delay and
lens mass modelings have been producing reasonable Hubble constants while using a
wrong time delay formula, it warrants special scrutinies of the fidelity of the analyses.

\subsection{The Isotropic Schwarzschild Metric}

The so-called isotropic Schwarzschild metric is of the following form, 
\opeqn
 ds^2 = - G(u) d\eta^2 + F(u) (du^2 + u^2 d\Omega^2) ,
 \label{eqSchwarzschildIso}
\cleqn
and can be obtained from 
the standard Schwarzschild metric by a coordinate transformation: $r \mapsto u$.
\opeqn
u = \dfrac{1}{2}\left( r- r_s/2 + (r^2- r r_s)^{1/2}\right);
\label{eqRadialIso}
\cleqn
\opeqn
F(u) = \dfrac{(u+a)^4}{u^4} ; \quad G(u) = \dfrac{(u^2-a^2)^2}{(u+a)^4}; \quad a \equiv \dfrac{r_s}{4}
\cleqn
The equations of motion are
\opeqn
 \dfrac{d\eta}{dp} = \dfrac{1}{G} ; \ \ \
 \dfrac{d\psi}{dp} = \dfrac{J}{r^2F}; \ \ \
 F \left(\dfrac{dr}{dp}\right)^2 = \dfrac{1}{G} - \dfrac{J^2}{r^2 F}.
\cleqn
The deflection angle and the flight time of a photon trajectory are
\opeqn
\int d\psi = \left( \dint_{u_2}^{u_0} +\dint_{u_0}^{u_1} \right ) 
           \dfrac{1/u \ \ du }{ \sqrt{ \dfrac{FG_0 u^2}{G F_0 u_0^2}-1 }  };
\label{eqAngleIntegralIso}
\cleqn
\opeqn
\int d\eta = \left( \dint_{u_2}^{u_0} +\dint_{u_0}^{u_1} \right ) 
           \dfrac{\sqrt{F/G} \ \ du }{ \sqrt{ 1-\dfrac{G F_0 u_0^2}{F G_0 u^2} }  }.
\label{eqTimeIntegralIso}
\cleqn
If one examines the deflection angle integral (\ref{eqAngleIntegral}) with the factor
$r_s$ in mind, half the deflection angle comes from the term $A$ in the numerator
and the other half comes from the $B$ term in the denominator.  $F = G^{-1}$ in the
linear order, and one can see immediately that eq.(\ref{eqAngleIntegralIso}) produces
the same deflection angle as eq.(\ref{eqAngleIntegral}).
If we compare the time integrals eq.(\ref{eqTimeIntegral}) and eq.(\ref{eqTimeIntegralIso}),
the latter has the combined factor $G/F\approx G^2$ instead of $B$ in the denominator 
and the extra factor doubles the third term in eq.(\ref{eqHalfTime}) resulting in
\begin{eqnarray}
\eta(u, u_0)  = 
 \sqrt{u^2-u_0^2}  
+ 2 GM \ln \left(\dfrac{u+\sqrt{u^2-u_0^2}}{u_0}\right) \cr
          + 2 GM\left(\dfrac{u-u_0}{u+u_0}\right)^{1/2} + {\cal O}({r_s^2}/{u_0^2}).
\label{eqHalfTimeIso}
\end{eqnarray}
In fact, $\eta(u,u_0)$ can be obtained from $t(r,r_0)$ by using eq.(\ref{eqRadialIso}).
In the linear order in $r_s$, $u = r - r_s/2$, and the extra factor in the third term in
eq.(\ref{eqHalfTimeIso}) can be seen coming from the first term in eq.(\ref{eqHalfTime}).
\opeqn
(r^2-r_0^2)^{1/2} = (u^2-u_0^2)^{1/2} + \dfrac{r_s}{2}\left(\dfrac{u-u_0}{u+u_0}\right)^{1/2}.
\label{eqRadialExtra}
\cleqn
So we can say that it is algebraically consistent, but it leaves a question as to 
which is the time the observer's clock will be measuring. As for the arrival time
difference between two gravitationally lensed images, the third term is irrelevant
in the linear order because it is effectively a constant.
\opeqn
 2 GM\left(\dfrac{u-u_0}{u+u_0}\right)^{1/2} = 2GM \left( 1 - \dfrac{u_0}{u}\right)
 = 2GM
\cleqn
where the expansion is made in $u_0/u << 1$. 

Incidentally, it is worth noting that $r_0/r$ is of the same order of $r_s/r_0$ for bright images.
They are near the Einstein ring, and hence
$r_0^2 \approx R_E^2 = 4GMD \approx 2 r_s r$ (see eq.(\ref{eqSingle})).   
In other words, $r_s/r = {\cal O}({r_s^2}/{r_0^2})$ for bright images.
As for the gravitational lensing by the Sun as observed from the Earth, the Einstein 
ring radius is much smaller than the size of the Sun ($R_E = R_\odot/23.2$), and the
only visible ``image" is highly unmagnified and $1 >> r_0/r >> r_s/r_0$.

Note that $(r^2-r_0^2)^{1/2}$ in eq.(\ref{eqRadialExtra})
 is the distance of the straight path the photon would have flown
when $M=0$. So, let's reexamine eq.(\ref{eqAngleIntegral}) to see if the angular
span of the straight path generates an $r_s$-dependent extra term upon the 
coordinate transformation.
\begin{eqnarray*}
\Delta\phi (r,r_0) =  \dfrac{\pi}{2} - \sin^{-1}\left(\dfrac{r_0}{r}\right) + \dfrac{r_s}{r_0}
 +  {\cal O}({r_s^2}/{r_0^2}),
\end{eqnarray*}
The first two terms are the angular span of the straight path from $r$ to $r_0$,
and the second term does generate an extra term.
In the linear order in $r_s$,
\opeqn
 \sin^{-1}\left(\dfrac{r_0}{r}\right) = \dfrac{r_0}{r} = \dfrac{u_0}{u} + \dfrac{r_s(u-u_0)}{2u^2}.
 \label{eqAngleExtra}
\cleqn
It is then curious because the extra term doesn't show up in the first order integration 
of eq.(\ref{eqAngleIntegralIso}).
\opeqn
\Delta\psi (u,u_0) =  \dfrac{\pi}{2} - \sin^{-1}\left(\dfrac{u_0}{u}\right) + \dfrac{r_s}{u_0}
 +  {\cal O}({r_s^2}/{u_0^2}).
\cleqn
We can reason that it is consistent because the extra term is effectively of a second order.
\opeqn
\dfrac{r_s(u-u_0)}{2u^2} \approx \dfrac{r_s}{2u} = {\cal O}({r_s^2}/{u_0^2}).
\cleqn
We will see in a later section that the second order perturbation in $r_s$ does
generate terms of the form $r_s^2/r_0^2$ which is of the order of $r_s/r$ for bright
images.

Upon looking at eq.(\ref{eqHalfTimeIso}), we may
reason that the extra time is due to that $u$ ($u < r$) measures a path that is 
closer to the mass where the potential is deeper and so the photon speed is smaller.
However, it still remains a question which formula for an observer to use between 
eq.(\ref{eqHalfTime}) and eq.(\ref{eqHalfTimeIso}) to compare with its clock time.
(The difference between the ``observer"'s coordinate time and proper time is the 
same in the linear order in the two coordinate systems as one can see in the 
fourth term of eq.(\ref{eqProperTime}).) Given a mass $M$, which is rightfully considered
a coordinate-invariant, we can imagine an abstract space and two abstract points $S$ and $O$
on it and a null geodesic connecting the two points. A coordinate system is a means to 
describe the abstract situation. What we are at a loss seems to be whether we are 
identifying the same abstract points $S$ and $O$ in the two well-known coordinate systems 
in which we examined the time delays and deflection angles. It should be worth
sorting it out unambiguously.

\section{Geometric Path Difference Term and Gravitational Potential Term}

Let's write the isotropic Schwarzschild metric in eq.(\ref{eqSchwarzschildIso})
in the linear approximation as CK75 did for more general gravitational fields.
\opeqn
 ds^2 = -(1-\dfrac{r_s}{u}) d\eta^2 + (1+\dfrac{r_s}{u}) (du^2 + u^2 d\Omega^2).
\cleqn
(CK75 uses (two times) the Newtonian potential $-2\Phi(u)$ instead of $r_s/u$.)
For a photon trajectory, $ds^2=0$, and the time integral can be written formally
as follows.
\opeqn
 \int d\eta = \int ds_3 + \int \dfrac{r_s}{u} ds_3,
 \label{eqEtaCK}
\cleqn
where $ds_3$ denotes the length element of the photon 
  trajectory in 3-space (to distinguish from the four dimensional line element $ds$).
   CK75 interpreted,
\begin{verse}
``The first term is the time due to the length of the path traveled and {\it must be computed to first
order in G} [our highlight]. The second is due to the
potential well through which the photon traveled."
\end{verse}
It is intuitive and reasonable. 
As far as we know, this is the origin of the current standard notion of the time delay
being constituted of two pieces: path difference and gravitational potential.

The question is what $ds_3$ is, and  we need to solve
for the trajectory to get the accurate coordinate time delay. 
It is  instructive to consider the straight path.
 If $ds_3$ is the line length of the straight path, then
$\int ds_3 = (r^2-r_0^2)^{1/2}$ and hence $ds_3 = (r^2-r_0^2)^{-1/2} r dr$; 
the integral eq.(\ref{eqEtaCK}) reproduces the first two terms of eq.(\ref{eqHalfTimeIso}).
The third term of eq.(\ref{eqHalfTimeIso}), which we discussed to be coordinate dependent
and irrelevant to the time delay between two images, is not recovered. 
For the Schwarzschild metric, we we have solved the equations of motion and 
know that the correct line element  is  $dp$.  It reproduces
the time integral eq.(\ref{eqTimeIntegralIso} exactly in the linear order. 
The integral eq.(\ref{eqEtaCK}) can be rewritten as follows in the linear order.
\opeqn
 \int d\eta =  \int dp + \int \dfrac{r_s}{u} d\ell = \int d\ell 
 + 2r_s\left(\dfrac{u-u_0}{u+u_0} \right)^{1/2} + \int  \dfrac{r_s}{u} d\ell .
\cleqn 
 where $\ell$ is the line parameter of the straight path. We can see that
 $\int dp$ generates  the geometric path difference effect ($\approx 2 r_s$) we have seen in 
 section \ref{sec:simply} and CK75 might have liked to have. 
 
 As we have repeatedly 
 stated, the path difference term $2r_s$ cancels out for the arrival time difference of 
 two images for a Schwarzchild lens. Thus for a general potential where the form of
 the correct line element $dp$ may not be easily obtained, it may be reasonable to 
 use the straight line element.
 
 In section \ref{sec:simply}, we ``tried to do better" and obtained the ``inverse quadratic
term" which we know to cancel out the true time delay. It is not any better than CK75's
doubling. As we have been
repeating it, the magnitude of the ``inverse quadratic term" is approximately $2r_s$.
The fault seems to be lie in that we introduced the second order in angles 
while the premise is the linear order small angle approximation. CK75 erred the same. 
The geometric relation
$r_s/r_0 \approx r_0/r$, where r is the position of the source or observer, 
seems to be the source of ready confusion.
The lesson may be that one needs to be as systematic as possible.
There is no obvious way to obtain the correct line element $dp$ without solving the
equations of motion, and the best bet is the straight line element. It is unlikely that
CK75 would have been able to get the correct geometric path difference. 
However, the notion of geometric time difference is fine.

\section{The Second Order Corrections}

The second order terms of the flight time and angular span are presented here.
If we define 
\opeqn
 \epsilon \equiv \dfrac{r_s}{r} ; \quad  \epsilon_0 \equiv \dfrac{r_s}{r_0}; \quad \xi \equiv \dfrac{r_0}{r},
\cleqn
the part of the time integral eq.(\ref{eqTimeIntegral}) that is of the second order in $r_s$ is
\begin{eqnarray}
\Delta t_2(r,r_0) = \epsilon_0^2 \dint  \dfrac{3 \xi^2}{2(1-\xi^2)^{1/2}}
     +\dfrac{3 \xi^4 (1-\xi)^2}{8(1-\xi^2)^{5/2}}  dr      \\
     = 
        \dfrac{r_s^2}{r_0} 
     \left[ \dfrac{15}{8} (\dfrac{\pi}{2}-\sin^{-1}\xi )
                +  \dfrac{(1-\xi)}{4(1-\xi^2)^{2/3}} -\dfrac{6-5\xi}{8(1-\xi^2)^{1/2}} 
      \right] .
\end{eqnarray}
If we set $r=\infty$, $\Delta t_2(r=\infty,r_0) = 2.445\, r_s^2/r_0$. It is good to find the
coefficient of the second order to be of ${\cal O}(1)$.
The second order part of the deflection angle eq.(\ref{eqAngleIntegral}) is
\opeqn
\Delta \phi_2(r,r_0) 
 = \epsilon_0^2
 \dint \dfrac{\xi^2}{r_0} \left[ \dfrac{3 \xi^2}{8(1-\xi^2)^{1/2}}
      + \dfrac{3\xi^2(1-\xi)}{4(1-\xi^2)^{3/2}}
     +\dfrac{3 \xi^4 (1-\xi)^2}{8(1-\xi^2)^{5/2}} \right] dr 
\cleqn
\opeqn
     = 
     \dfrac{r_s^2}{r_0^2} 
     \left[  \dfrac{15}{16} (\dfrac{\pi}{2}-\sin^{-1}\xi )
                +  \dfrac{(1-\xi)}{4(1-\xi^2)^{2/3}} -\dfrac{6-5\xi}{8(1-\xi^2)^{1/2}} 
                +\dfrac{3\xi(1-\xi^2)^{1/2}}{16}       \right] .
\cleqn

\section{Conclusion}
\label{sec:conclusion}

We examined the current standard time delay formula for a Schwarzschild black hole lens
in the weak field (or small angle) approximation. We find that the current standard time delay 
formula is wrong. It effectively doubles the true time delay. The `quadratic term"
is the result of a wrong calculation of the geometric path difference. 

The absence of the ``quadratic term" thwarts the 
claim of the relation between the time delay and the lens equation through the Fermat's
principle. However, the CSTD formula can be considered a 
generating function of the lens equations because it works, even though it is unclear
presently what significance it has. 

We call to reanalyze the strong gravitational lenses with time delay measurements. 
We expect the correct time delay formula to help for better lens modelings. On the other hand,
the current practices of reasonably good fittings to the Hubble constant while using a 
wrong time delay formula makes one worried lest the data analyses tend to converge on 
desirable values. We call for exceptional scrutinies of the fidelity of the analysis methods.

We identified a term in time delay that is dependent 
on the coordinate systems, even though it is benign for the arrival time difference 
of two images. The coordinate dependence of the ``measurables" 
seems to require some careful thoughts as to what is what we measure.

\bsp

\label{lastpage}

\end{document}